\documentclass[twocolumn,english]{revtex4}
\usepackage{times}
\usepackage[T1]{fontenc}
\usepackage[latin1]{inputenc}
\usepackage{graphicx}
\usepackage{amssymb}

\makeatletter

\usepackage{babel}
\makeatother
\begin{document}

\title{Degenerate Fermi gas in a combined harmonic-lattice potential}

\author{P. B. Blakie$^{1}$, A. Bezett$^{1}$, P. Buonsante$^{2,1}$}

\affiliation{$^{1}$Jack Dodd Centre for Photonics and Ultra-Cold Atoms, Department
of Physics, University of Otago, P.O. Box 56, Dunedin, New Zealand}

\affiliation{$^{2}$ Dipartimento di Fisica, Politecnico di Torino, Corso Duca
degli Abruzzi 24, I-10129 Torino, Italy.}

\begin{abstract}
In this paper we derive an analytic approximation to the density of
states for atoms in a combined optical lattice and harmonic trap potential
as used in current experiments with quantum degenerate gases. We compare
this analytic density of states to numerical solutions and demonstrate
its validity regime. Our work explicitly considers the role of higher
bands and when they are important in quantitative analysis of this
system. Applying our density of states to a degenerate Fermi gas we
consider how adiabatic loading from a harmonic trap into the combined
harmonic-lattice potential affects the degeneracy temperature. Our
results suggest that occupation of excited bands during loading should lead to more
favourable conditions  for realizing degenerate Fermi gases in optical lattices.
\end{abstract}
\maketitle

\section{Introduction}

Tremendous progress has been made in the preparation, control and
manipulation of Fermi gases in the degenerate regime \cite{DeMarco1999a,Schreck2001a,O'Hara2002a,Modugno2002a,Gupta2003a,Regal2003a,Cubizolle2003a}.
Such systems have many potential applications in the controlled study
of fermionic superfluidity and the production of ultra-cold molecules.
Another area of developing theoretical interest is in the physics
of fermions in optical lattices \cite{Hofstetter2002,Rabl2003,Santos2004,Rigol2004a,viverit2004a},
and experiments have already begun to examine the properties of Fermi-gases
(prepared as boson-fermion mixtures) in one-dimensional \cite{Modugno2003a,Ott2004a}
and three-dimensional \cite{Stoferle2006a,Kohl2005a,Ospelkaus2006a}
optical lattices.

\begin{figure}[t]
\includegraphics[width=3.4in,keepaspectratio]{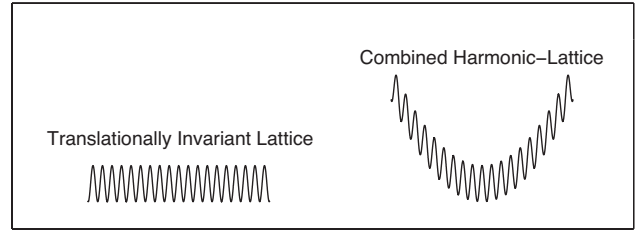}

\caption{\label{fig:CombLatt} Schematic diagram comparing the translationally
invariant lattice to the combined harmonic lattice considered in this
paper.}
\end{figure}

Optical lattices have many features in common with crystals where
a periodic lattice is also present, and many of the ideas and techniques
from solid state physics have been applied to this system. A unique
property of optical lattices, as realized in experiments, is that
the periodic lattice is accompanied by a harmonic confining potential,
arising from an external magnetic trap or from effects related to
the focused laser beams used to form the lattice. We shall refer to
this potential as the \emph{combined harmonic-lattice potential},
which is the main subject of the investigation presented in this paper
(see Fig. \ref{fig:CombLatt}). 

Whilst the harmonic potential and \emph{translationally invariant}
periodic potential are well characterized individually, their properties
when combined are not as well-understood. Even though the harmonic
trap is often much weaker than the confinement provided by each lattice
site, its effect on the spectrum and properties of the system can
hardly be considered small: it breaks the translational invariance
of the system and changes the nature of the energy states in the deep
lattice from compressed bands, to a set of unbounded overlapping bands.
Several recent articles have considered aspects of this system \cite{Hooley2004a,viverit2004a,Rigol2004b,Rey2005a}.
In the context of a tightbinding model of ultra-cold bosons, the spectrum
of the combined potential appears to have been first considered by Polkovnikov \emph{et al}. \cite{Polkovnikov2002a}. Refs. \cite{Hooley2004a,Rigol2004b}
have made detailed studies of the combined potential spectrum (also within
a tightbinding description), and closed-form solutions to this problem were recently given by Rey \emph{et al}. \cite{Rey2005a}. In Refs. \cite{viverit2004a,Ruuska2004a}
 an ideal gas of fermions in a 1D combined potential was examined without
making the tightbinding approximation. All of these studies have confirmed
that, for appropriate parameter regimes, parts of the single particle
spectrum will contain localized states. This is in contrast to the translationally
invariant system, where inter-atomic interactions or disorder are
needed for localization to occur (e.g. see \cite{Jaksch1998a,Buonsante2006a}).
In the combined potential localization arises solely from single-particle
effects. Experiments with ultra-cold (though non-condensed) bosons
\cite{Ott2004b} have provided evidence for these localized states.

Many of the physical phenomena that are suitable to experimental investigation
in optical lattices are sensitive to temperature and it is therefore
of great interest to understand how the temperature of a quantum degenerate
gas changes with lattice depth. Experimental results by Kastberg \emph{et
al.} \cite{Kastberg1995a} in 1995 showed that loading laser cooled
atoms into a three-dimensional optical lattice caused the atoms to
increase their temperature %
\footnote{In fact this study used adiabatic de-loading to reduce the temperature
of the constituent atoms.%
}. In previous work we have studied how the depth of a translationally
invariant lattice affects the thermodynamic properties of quantum
degenerate Bose \cite{Blakie2004b} and Fermi \cite{Blakie2005a}
systems (also see \cite{Werner2005a,Rey2006c,Ho2007a,PBlakie2007a}). The most important predictions of those studies relate to
the temperature changes induced by the lattice, in particular that
in appropriate regimes increasing the lattice depth could be used
to cool the system. For Fermi gases this cooling effect was used to
predict that loading into an optical lattice could be used to enhance
the conditions for observing the superfluid transition \cite{Hofstetter2002}.
However, recent work with a tightbinding model has shown that including
the effects of the harmonic potential leads to an increase in the temperature
of the system (compared to the Fermi temperature) by
a factor of 2 \cite{Kohl2006a}, demonstrating the importance of fully
treating the combined potential. 

In this paper we consider the thermal properties of a Fermi gas in
the combined harmonic-lattice potential including the role of higher
vibrational states. We derive an analytic approximation to the single-particle
density of states and compare this analytic density of states to numerical
solutions and determine validity criteria. We give characteristic energy
and atom-number scales which can be used to predict when higher band
effects will be important.  Applying these density
of states to a degenerate Fermi gas we consider how adiabatic loading
from a harmonic trap into the combined harmonic-lattice potential
affects the degeneracy temperature over a wide parameter regime.
Our results show that beyond tightbinding effects provide an appreciable correction to the adiabatic loading calculations presented in Ref. \cite{Kohl2006a}. Also, in the regime where higher band states are appreciably occupied, we find that the degeneracy temperature tends to increase to a lesser extent or even reduce  during loading, suggesting that this regime might be useful for preparing a strongly degenerate Fermi gas in an optical lattice. Recently molecule production was used to measure the temperature of fermions in an optical lattice \cite{Stoferle2006a}  and using this (or other approaches \footnote{E.g. Bragg and Raman spectroscopy is sensitive to temperature and fluctuation effects for bosons in optical lattices \cite{Rey2005b,Blakie2006c} we expect that generalizations of these probing schemes to fermions will also be useful for measuring temperature.}) the predictions of our work can be examined in current experiments.

\section{Formalism}

\subsection{Analytic density of states in the inhomogeneous lattice potential }

Here we consider the properties of a system described by the single
particle Hamiltonian\begin{equation}
H=-\frac{\hbar^{2}}{2m}\nabla^{2}+V_{c},\label{eq:Hsp}\end{equation}
where the \textit{combined} potential $V_{c}$ is formed by an optical
lattice potential and harmonic trap potential, i.e.\begin{eqnarray}
V_{{\rm c}} & = & V_{h}+V_{l},\label{eq:VharmLatt}\\
V_{h} & = & \frac{1}{2}m(\omega_{x}^{2}x^{2}+\omega_{y}^{2}y^{2}+\omega_{z}^{2}z^{2}),\label{eq:Vharm}\\
V_{l} & = & V_{0}[\sin^{2}(kx)+\sin^{2}(ky)+\sin^{2}(kz)].\label{eq:VLatt}\end{eqnarray}
The harmonic trap is taken to be anisotropic, with angular frequencies
$\{\omega_{x},\omega_{y},\omega_{z}\}$ along the coordinate directions.
The lattice is of depth $V_{0}$, $k$ is the wavevector of the counter-propagating
light fields used to form the lattice and $a=\pi/k$ is the direct
lattice vector. We also use $k$ to define the recoil frequency $\omega_{R}=\hbar k^{2}/2m$,
and associated recoil energy $E_{R}=\hbar\omega_{R}$. The combined
harmonic-lattice potential, as defined in Eqs. (\ref{eq:VharmLatt})-(\ref{eq:VLatt})
has minima (i.e. lattice sites) at positions $\mathbf{r}=n_{x}a\hat{\mathbf{x}}+n_{y}a\hat{\mathbf{y}}+n_{z}a\hat{\mathbf{z}},$
where $\{\hat{\mathbf{x}},\hat{\mathbf{y}},\hat{\mathbf{z}}\}$ are
unit vectors, and $\{ n_{x},n_{y},n_{z}\}$ are integers that are
convenient for labelling particular lattice sites. We note that our
highly symmetric choice of the lattice potential, having a site coincident
with the minimum of the harmonic potential, may be rather difficult
to arrange experimentally. However, our primary interest is in the
thermodynamic properties of the system which are insensitive to this
symmetry.

We are interested in the limit where the lattice dominates the short-length
scale properties of the system, and will take $a\ll a_{{\rm ho}}$,
where $a_{{\rm ho}}=\min\{\sqrt{\hbar/m\omega_{j}}\}_{j=x,y,z}$ is
the smallest harmonic oscillator length. This removes our need to
consider systems where extremely tight harmonic confinement causes
all the atoms to coalesce to a single site.

\subsubsection{1D spectrum}

Viverit \emph{et al}. \cite{viverit2004a} have shown for the one-dimensional
case of Eq. (\ref{eq:Hsp}) that when the lattice is sufficiently
deep the eigenstates are localized to lattice sites. For our symmetric
potential the eigenstate localized at site $n$ will necessarily also
localize at site $-n$, so strictly we should not call such states
localized. However, this property is fragile to any asymmetry in the
system, and has no discernible effect on the energy spectrum (which
is of primary interest to us). In this regime, the lattice site index
$n$ forms a convenient quantum number for the eigenstates, specifying
the site where the state is localized. The respective energy eigenvalue
is given by the value of the harmonic potential at that site, i.e.
\begin{equation}
\epsilon_{n}^{(0)}=\frac{1}{2}ma^{2}\omega^{2}n^{2},\label{eq:1Dgbspec}\end{equation}
where  $\omega$ is the trap frequency and we have set to zero the
zero-point energy associated with the confinement in each lattice
site. We will refer to these states as \textit{ground band} states
for clarity. The condition for localization is $\Delta E(n)>J_{0}$,
where $\Delta E(n)\equiv\epsilon_{n+1}^{(0)}-\epsilon_{n}^{(0)}$
is the difference in (harmonic trap) potential energy between lattice
site $n$ and $n+1$, and $J_{0}$ is the tunneling between sites %
\footnote{We determine $J_{0}$ using band structure calculations, see \cite{Blakie2004a}%
}. This requirement is most difficult to satisfy near the trap potential
minimum, where the difference in energy between adjacent sites is
least. This validity condition is equivalent to \begin{equation}
|n|>n_{{\rm crit}}\equiv q/8,\label{eq:ncrit}\end{equation}
where $n_{{\rm crit}}$ is the approximately the index of the closest
site to the origin for which the localization condition is satisfied
and $q\equiv8J_{0}/m\omega^{2}a^{2}$ is a dimensionless parameter
which we discuss below (originally defined in \cite{Rey2005a}).

While states satisfying Eq. (\ref{eq:1Dgbspec}) are valid for sufficiently
large values of $n,$ this expression neglects the existence of excited
vibrational states, which for the case of a translationally invariant
lattice would correspond to the first excited band. The energy scale
for the emergence of these excitations is $\epsilon_{\rm{gap}}$, and an
analytic approximation for this quantity is calculated in Appendix
\ref{sec:Egap}. Like the ground band states, these states will also
localize when the difference between potential energy at neighbouring
sites exceeds the tunneling matrix element for the first excited band,
which we denote $J_{1}$. Where this condition is satisfied the spectrum
of these states will take the analytic form\begin{equation}
\epsilon_{n}^{(1)}=\frac{1}{2}ma^{2}\omega^{2}n^{2}+\epsilon_{\rm{gap}}.\label{eq:1Dexbspec}\end{equation}
For clarity we shall refer to these as the \emph{first excited
band} states. This argument could be extended to additional bands
of states, in particular, at an energy of roughly $2\epsilon_{\rm{gap}}$
the next vibrational states will be accessible. However, the tunneling
rate $J_{m}$ increases with the band index $m$, making the localization
condition more difficult to satisfy, and analytic estimates for the
energy gap to higher bands less accurate. Indeed, for single particle
energies large compared to the lattice depth the spectrum will cross
over to that of a harmonic oscillator. Additionally, for typical experimental
parameters, the first two bands include sufficiently many states, and a large enough energy range to provide an accurate description  description of the system. Of course, high temperatures, or large lattice constants would require consideration of additional bands.

\subsubsection{3D spectrum and density of states}

For more than one spatial dimension the wavefunction localization
may be broken by neighbouring sites which have approximately the same
local potential value. However since Eq. (\ref{eq:Hsp}) is separable,
the 3D ground band spectrum is completely determined by the one dimensional
results (\ref{eq:1Dgbspec}) and (\ref{eq:1Dexbspec}), and under
the assumption of localized states (see below), the ground and first
excited band spectra are given by the expressions\begin{eqnarray}
\epsilon_{n_{x}n_{y}n_{z}}^{(0)} & = & \frac{1}{2}ma^{2}(\omega_{x}^{2}n_{x}^{2}+\omega_{y}^{2}n_{y}^{2}+\omega_{z}^{2}n_{z}^{2}),\label{eq:localspectrum}\\
\epsilon_{n_{x}n_{y}n_{z}}^{(1)} & = & \frac{1}{2}ma^{2}(\omega_{x}^{2}n_{x}^{2}+\omega_{y}^{2}n_{y}^{2}+\omega_{z}^{2}n_{z}^{2})+\epsilon_{\rm{gap}},\label{eq:localspectrum2}\end{eqnarray}
respectively. Because of the lattice symmetry, there are 3 equivalent (and completely overlapping)  first excited bands. Thus each energy state specified by the quantum numbers $\{n_x,n_y,n_z\}$ in (\ref{eq:localspectrum2}) is three-fold degenerate \footnote{This is in addition to the degeneracy arising from spatially equivalent lattice sites, e.g. for $\{n_x,n_y,n_z\}$ all non-zero there are 8 states of the same energy found by taking $\{\pm n_x,\pm n_y,\pm n_z\}$.} . This degeneracy
would be broken if the lattice depth was different in each direction causing the first excited bands to separate in energy,
but we do not consider that case here. The separability of the potential
means that the validity conditions discussed for the one-dimensional
case apply immediately, using the respective trap frequency in each
direction, i.e. $n_{j}>n_{{\rm crit}}^{(j)}\quad j=x,y,z$, where
$n_{{\rm crit}}^{(j)}$ is $n_{{\rm crit}}$ evaluated according to
(\ref{eq:ncrit}) using the trap frequency along direction-$j$.$ $

The density of states for the spectra in Eqs. (\ref{eq:localspectrum})
and (\ref{eq:localspectrum2}) is given by
\begin{eqnarray}
g_{c}(\epsilon) & = & \frac{16}{\pi^{2}}\left(\frac{\omega_{R}}{\bar{\omega}}\right)^{3/2}\frac{\epsilon^{1/2}}{(\hbar\bar{\omega})^{3/2}}\label{eq:AnalyticInHomogDOS}\\
 &  & +\frac{48}{\pi^{2}}\left(\frac{\omega_{R}}{\bar{\omega}}\right)^{3/2}\frac{(\epsilon-\epsilon_{\rm{gap}})^{1/2}}{(\hbar\bar{\omega})^{3/2}}\theta(\epsilon-\epsilon_{{\rm gap}}),\nonumber \end{eqnarray}
where $\theta(x)$ is the unit step function, and $\bar{\omega}\equiv\sqrt[3]{\omega_{x}\omega_{y}\omega_{z}}$.
The first term, corresponding to the ground band contribution to the
density of states, exhibits a $\sqrt{\epsilon}$-scaling with energy,
similar to that of a homogeneous gas of free particles \cite{Kohl2006a}.
The second term includes the contribution of the first excited band
states that occur at energies $\epsilon>\epsilon_{\rm{gap}}$. As was noted
in our discussion of the validity conditions of the spectra, additional
bands will become accessible at $\epsilon\sim2\epsilon_{\rm{gap}},$ and
so Eq. (\ref{eq:AnalyticInHomogDOS}) should only be used in situations
where $k_{B}T$ and the Fermi energy are much less than $2\epsilon_{\rm{gap}}.$

\subsection{Numerical Results for single particle spectrum}

\subsubsection{1D spectrum}

\begin{figure}[t]
\includegraphics[width=3.4in,keepaspectratio]{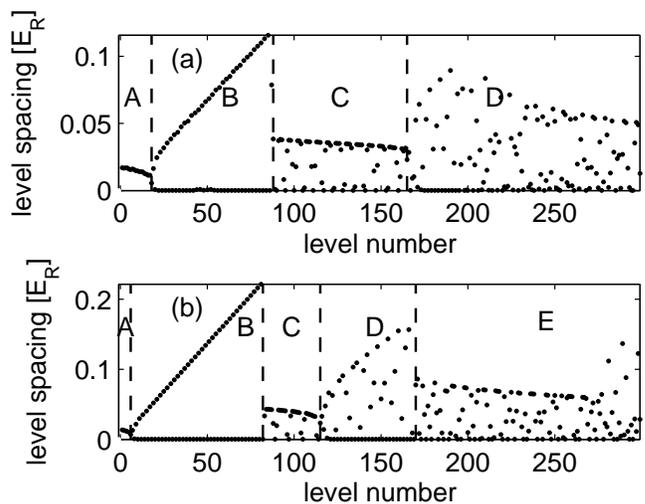}
\caption{\label{fig:ExptSpec} Energy level spacing for the 1D eigenspectrum
relevant to the experimental setup in Ref. \cite{Kohl2005a}. (a)
$V_{0}=5E_{R}$, $\omega=0.024\omega_{R}$ (b) $V_{0}=10E_{R}$, $\omega=0.033\omega_{R}$.
Parameters derived for $^{40}K$ with lattice made from counter propagating
$\lambda=826nm$ lasers with harmonic confinement arising from the
focused lasers (beam waist taken to be $50\mu m$). Labels A-E explained
in text.}
\end{figure}

Here we show some typical results of the 1D spectrum of the combined
potential in Fig. \ref{fig:ExptSpec}. In Refs. \cite{Hooley2004a,Rigol2004b,Rey2005a,Polkovnikov2002a}
a detailed analysis of the spectrum has also been made, but in the
tight binding (Hubbard) limit where only the vibrational ground state
of each lattice site are included. Those studies considered a wide
parameter regime of trapping frequencies and lattice depths, however
in the tight-binding limit a single parameter describing the ratio
of tunneling to harmonic potential is sufficient to characterize the
nature of the eigenspectrum. Several choices of  parameter
are used in the literature, and we follow the choice of Rey \emph{et al}.
\cite{Rey2005a} who define $q\equiv4J_{0}/(\frac{1}{2}m\omega^{2}a^{2}).$
With the inclusion of higher bands this single parameter by itself  is insufficient
to characterize the spectrum and both the lattice depth and harmonic
confinement parameters are independently important. %
 In Fig. \ref{fig:ExptSpec} we show the level spacing obtained from
numerical diagonalization of the one-dimensional case of Eq. (\ref{eq:Hsp}).
The parameters for this calculation were taken to correspond to those
of the experiment \cite{Kohl2005a}. For Fig. \ref{fig:ExptSpec}(a)
with $V_{0}=5E_{R}$ we find that $q\approx193$ and for Fig. \ref{fig:ExptSpec}(b)
with $V_{0}=10E_{R}$ we find that $q\approx28$. It is useful to
compare our results to Fig. 1(b) of Ref. \cite{Rigol2004b} to assess
the effects of higher band states on the spectrum. We have segmented
our spectrum with vertical dashed lines and indicated the characteristic
regions by the letters A-E, which we explain: In region A tunneling
dominates over the offset between lattice sites and the eigenstates
are delocalized (for these states Eq. (\ref{eq:1Dgbspec}) will be
a poor approximation). In region B localized states emerge which are
approximately degenerate in energy (the spacing between every second
eigenvalue is approximately zero). Because of the localizations (and
applicability of Eq. (\ref{eq:1Dgbspec})), the energy spacing between
degenerate pairs increases linearly throughout this region. Regions
A and B, as identified in Fig. \ref{fig:ExptSpec}, are equivalent
to those in Fig. 1(b) of Ref. \cite{Rigol2004b}, and in general we
find that the tightbinding description is in good qualitative agreement
with our more general analysis. For the next regions (i.e. C, D etc.)
the role of higher bands is essential and cannot be described within
a simple tightbinding model. Region C has a similar appearance to
region A, and consists of delocalized excited band states. The scatter
in level spacing seen in C occurs because ground band states are also
available in this energy range. In region D the excited bands state
have become localized (similar to what happened in region B to the ground band states). In region E we have reached a sufficiently
high energy scale that a 3rd band of states have begun to contribute. 

The harmonic confinement used in Fig. \ref{fig:ExptSpec} originates
from the dipole conferment provided by the focused lasers used to
make the lattice%
\footnote{The harmonic frequency is given by $\omega=\sqrt{8V_{0}/mw^{2}}$
where $w$ is the beam waist (also see \cite{Kohl2006a}). %
}, and as the lattice depth increases so does the strength of harmonic
confinement. We note that for Fig. 1(b) of \cite{Rigol2004b} $q\approx13\times10^{6}$
which is many orders of magnitude away from that found in current
experiments %
\footnote{We calculate that $q=13\times10^{6}$ roughly corresponds to a $V_{0}=4E_{R}$
deep lattice and harmonic confinement of oscillator frequency $0.8$
Hz. %
}. The number of states in region A roughly scales as $\sqrt{q}$ \cite{Rey2005a}
so that in experimentally realized lattices region A is quite small
and the majority of ground band states are well-localized.

\subsubsection{Critical site index}

\begin{figure}[t]
\includegraphics[width=3.5in,keepaspectratio]{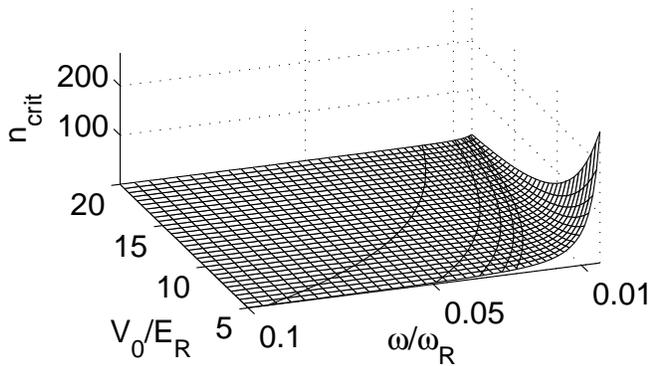}
\caption{\label{cap:ncrit} The critical site number for localization, as
defined in Eq. (\ref{eq:ncrit}), as a function of lattice depth and
harmonic trap frequency. Contour lines are shown for $n_{{\rm crit}}=1,3,5$
and 7. }
\end{figure}
In Fig. \ref{cap:ncrit} we present $n_{{\rm crit}}$ (\ref{eq:ncrit})
for a wide range of lattice depths and trap frequencies (note that
we have used band structure calculations in a translationally invariant
lattice to determine $J_{0}$ for each value of $V_{0}$). We observe
that $n_{{\rm crit}}$ is large for small trap frequencies and shallow
lattices so that only neighboring lattice sites quite far from the harmonic trap
minimum have a sufficiently large potential difference to tunneling
ratio to cause eigenstate localization. In such cases the analytic
approximation for the eigenspectra given in Eqs. (\ref{eq:localspectrum})
and (\ref{eq:localspectrum2}), and density of states given in Eq.
(\ref{eq:AnalyticInHomogDOS}) will not be valid, and the result of
the full numerical diagonalization will be necessary.

With increasing lattice depth the ground band tunneling matrix element
decreases and the potential difference to tunneling ratio increases.
Thus we find that $n_{{\rm crit}}$ decreases with increasing $V_{0}$.
Similarly, increasing the harmonic trap frequency also leads to a
decrease in $n_{{\rm crit}}$. The particular value of $n_{{\rm crit}}$
that justifies the use of our analytic density of states (\ref{eq:AnalyticInHomogDOS})
depends on the parameters of system under consideration. If the system
extends over $N_{s}$ lattice sites in each direction, then $n_{{\rm crit}}\ll N_{s}$
will be sufficient to ensure that the majority of the occupied states
are well described by the localized spectrum.

\subsubsection{Density of states}
Equation (\ref{eq:AnalyticInHomogDOS}) for the density of states
in the combined potential is one of the central results of this paper.
In this section we present numerical results to confirm the validity
regime of this expression. To do this we diagonalize Eq. (\ref{eq:Hsp})
to obtain the single particle eigen-spectrum $\{\epsilon_{j}\}$ for
various trap frequencies and lattice depths. For the purposes of comparison
to the analytic results, it is useful to construct a smoothed density
of states, defined as \begin{equation}
\bar{g}(\epsilon)=\frac{1}{2\Delta\epsilon}\int_{\epsilon-\Delta\epsilon}^{\epsilon+\Delta\epsilon}\sum_{j}\delta(\epsilon-\epsilon_{j}),\label{eq:smoothedDOS}\end{equation}
giving the average number of eigenstates with energy lying within
$\Delta\epsilon$ of $\epsilon$. 
\begin{figure}[!h]
\includegraphics[width=3.4in]{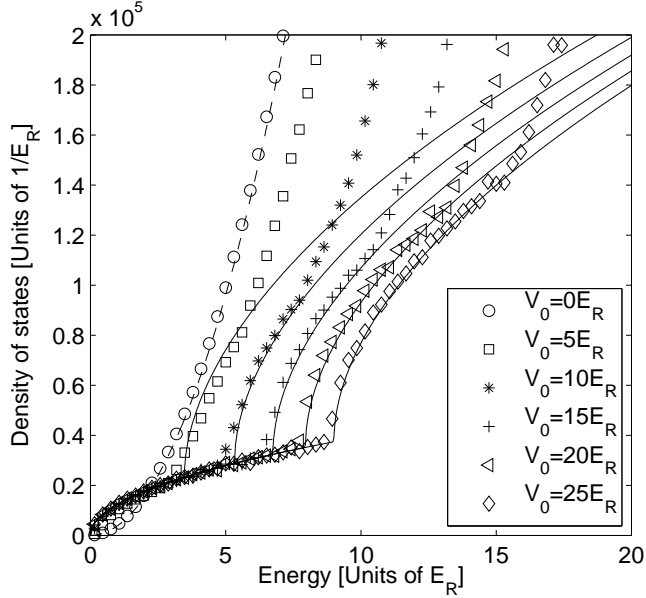}
\caption{\label{cap:DOSfig} Density of states for $\bar{\omega}=0.051\omega_{R}$.
Numerical results for the density of states are shown using various
markers (as labeled in the figure). The analytic density of states
given in Eq. (\ref{eq:AnalyticInHomogDOS}) is plotted for the cases
with $V\ge5E_{R}$ (solid line), using the expression $\epsilon_{\rm{gap}}=2\sqrt{V_{0}E_{R}}-E_{R}$
for the band gap energy. The harmonic oscillator density of states
is also shown (dashed line). The averaging interval used for the smoothed density of states is $\Delta\epsilon=0.303E_{R}.$}
\end{figure}

In Fig. \ref{cap:DOSfig} we compare the numerically calculated smoothed
density of states against the analytic result $g_{c}(\epsilon)$ for
various lattice depths in an isotropic trap of frequency $\bar\omega=0.051\omega_{R}$.
Agreement between the analytic and numerical calculations is seen
to improve as the lattice depth increases. We also observe that at
energies greater than approximately twice the gap energy the analytic
and numerical results begin to differ more significantly as the contribution
of additional bands become important (the gap energy in each case
is the energy at which the cusp in the analytic density of states
occurs).
\begin{figure}[!h]
 \includegraphics[width=3.4in]{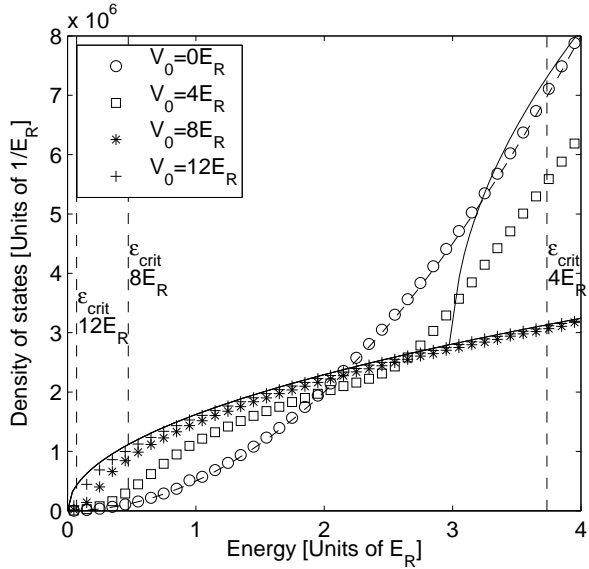}
\caption{\label{cap:detaileddos} The density of states for a combined harmonic
trap and lattice in the loose trap limit. Isotropic harmonic trap
with trap frequency $\bar\omega=0.01\omega_{R}$. The critical energy,
given by $\epsilon_{{\rm crit}}=m\bar\omega^{2}a^{2}n_{{\rm crit}}^{2}/2$,
is shown as a dashed vertical line and labeled by its corresponding
lattice depth. The averaging interval used for the smoothed density of states is $\Delta\epsilon=0.10E_{R}.$}
\end{figure}

It is of interest to more closely examine the reliability of the analytic
density of states at low energy scales and for weak harmonic traps.
In Fig. \ref{cap:detaileddos} we show such a comparison for an isotropic
trap of frequency $\bar\omega=0.01\omega_{R}$. For $V_{0}=4E_{R}$ the
analytic result is in poor agreement with the numerical result, as
the low energy states in this case are not localized. For $V_{0}=8E_{R}$
and more so $V_{0}=12E_{R}$, the analytic result is seen to provide
a useful description of the numerical density of states. To quantify
the degree of agreement seen in these results it is useful to consider
the energy of the (first) localized state at site $n\approx n_{{\rm {\rm crit}}}$,
i.e. $\epsilon_{{\rm crit}}=m\bar\omega^{2}a^{2}n_{{\rm crit}}^{2}/2,$
indicated as vertical dashed lines in Fig. \ref{cap:detaileddos}.
The quantity $\epsilon_{{\rm crit}}$ is the energy scale above which
the ground band states become localized, and hence the lowest energy
at which spectrum (\ref{eq:1Dgbspec}) is valid. For the case of $V_{0}=4E_{R},$
$\epsilon_{{\rm crit}}$ exceeds $\epsilon_{\rm{gap}}$ so that no
states in the ground band are localized below the energy scale at
which excited band states become accessible. For the case of $V_{0}=8E_{R}$,
$\epsilon_{{\rm crit}}\simeq0.5E_{R}$ and for energies above this
better agreement between the analytic and numerical results is observed
in Fig. \ref{cap:detaileddos}. For $V_{0}=12E_{R}$ the critical
energy is $\epsilon_{{\rm crit}}\simeq0.07E_{R}$ and the analytic
density of states furnishes better agreement at much lower energies.

In summary, we make the following observations about our analytic
expression for the density of states:
\begin{enumerate}
\item Within its regime of validity, the analytic density of states provides
an accurate description at intermediate energy scales, i.e. at energies
above $\epsilon_{{\rm crit}}$ where the spectrum is well-localized,
yet below $\epsilon\sim2\epsilon_{\rm{gap}}$ where additional bands become
accessible. 
\item The agreement between the analytic density of states and numerical
calculations improves with increasing lattice depth and increasing
trap frequency. 
\end{enumerate}

\section{Applications to Fermi gases }

\subsection{General Properties}

\subsubsection{Fermi energy}

The Fermi energy in the combined harmonic-lattice potential, $\epsilon_{F,c},$
is determined from the number of particles in the system according
to \begin{equation}
N=\int_{0}^{\epsilon_{F,c}}d\epsilon\, g_{c}(\epsilon).\label{eq:EFdefn}\end{equation}
To compute the Fermi energy from our analytic result (\ref{eq:AnalyticInHomogDOS})
it is convenient to define the \emph{cusp number} 
\footnote{So named, because the corresponding Fermi energy sits at the cusp
in the analytic density of states (e.g.. see Fig. \ref{cap:DOSfig}). %
} $N_{c}$ as the number of particles for which $\epsilon_{F,c}=\epsilon_{\rm{gap}}.$
Because the energy gap depends on the lattice depth, so does $N_{c}$.
Using Eqs. (\ref{eq:AnalyticInHomogDOS}) and (\ref{eq:EFdefn}) we
obtain \begin{equation}
N_{c}=\frac{32}{3\pi^{2}}\left(\frac{\omega_{R}}{\bar{\omega}}\right)^{3/2}\left(\frac{\epsilon_{\rm{gap}}}{\hbar\bar{\omega}}\right)^{3/2}.\label{eq:Nc}\end{equation}
For $N<N_{c}$ (i.e. $\epsilon<\epsilon_{\rm{gap}}$) only the first term
in the density of states (\ref{eq:AnalyticInHomogDOS}) is nonzero
in Eq. (\ref{eq:EFdefn})  and we can invert to obtain the Fermi energy\begin{equation}
\epsilon_{F,c}=\left(\frac{3\pi^{2}N}{32}\right)^{2/3}\frac{\hbar\bar{\omega}^{2}}{\omega_{R}},\qquad N<N_{c}.\label{eq:EFc1}\end{equation}
 For $N>N_{c}$ excited band states contribute. In this case a general
analytic expression for the Fermi energy in terms of $N$ is not available,
but for $(N-N_{c})\ll N_{c}$ we make a series expansion of the integral
of the density of states about $\epsilon_{\rm{gap}}.$ This expression
can then be solved perturbatively to yield the approximate expression\begin{equation}
\epsilon_{F,c}=\epsilon_{\rm{gap}}\left\{ 1+\delta-\frac{2\delta^{3/2}}{1+3\delta^{1/2}}\right\} ,\quad0<\delta\ll1\label{eq:EFc2}\end{equation}
where $\delta\equiv2(N-N_{c})/3N_{c}$ is the small parameter. 

It is also convenient to define the \emph{cusp depth} $V_{c}$, as
the lattice depth at which $\epsilon_{F,c}=\epsilon_{\rm{gap}}$. The cusp
depth is a function of the number of particles and harmonic trap frequency,
given by
\begin{equation}
V_{c}=\frac{\hbar\omega_{R}}{4}\left[\left(\frac{3\pi^{2}N}{32}\right)^{2/3}\left(\frac{\bar{\omega}}{\omega_{R}}\right)^{2}+1\right]^{2},\label{eq:Vc}\end{equation}
where we have made use of the analytic expression relating $\epsilon_{\rm{gap}}$ to $V_0$, as derived in Appendix \ref{sec:Egap}.

The two cusp parameters characterize the interesting features of the
combined harmonic-lattice system and can be interpreted as follows:
\begin{itemize}
\item $N_{c}$: For a system with fixed combined potential (i.e. fixed $V_{0}$,
$\{\omega_{j}\}$), $N_{c}$ is the maximum number of atoms that can
be accommodated in the ground band only. For $N>N_{c}$ the $T=0$
ground state of the system will contain excited band states. 
\item $V_{c}$: For a system with fixed atom number and harmonic trap frequencies,
$V_{c}$ is the smallest lattice depth for which the atoms can be
accommodated in the ground band. For $V_{0}<V_{c}$ the $T=0$ ground
state of the system will contain excited band states. 
\end{itemize}
These parameters motivate us to emphasize the distinctive properties of the energy spectrum in the combined potential as compared to the usual periodic lattice case. In the translationally
invariant lattice, there is a fixed number of single particle states
in each band (equal to the number of lattice sites), and for sufficiently
deep lattices (typically $V_{0}\gtrsim2E_{R}$) the ground and first
excited bands occupy disjoint energy regions separated by a finite
energy gap. In contrast, for the combined harmonic-lattice potential,
the energy bands are overlapping and can only be differentiated by
the local nodal structure of the wavefunctions at each site, where
they have a spatial character approximately given by harmonic oscillator
states (see Appendix \ref{sec:Egap}). This local structure of the
wavefunctions is apparent in experiment, and leads to states of different
bands (as we have defined them here) residing in distinctive regions
of momentum space in expansion images (e.g. see Refs. \cite{Kohl2005a,Denschlag2002a}).
If it is desirable to restrict the system to access only states of
the ground band, so as to realize a system well-described by a Hubbard
model, then according to our above prescription this necessarily requires  $N<N_{c}$ or equivalently $V_{0}>V_{c}$, in addition to having sufficiently low temperature.

\subsection{Sommerfeld analysis of isentropic loading of a degenerate Fermi gas}

The properties of a quantum degenerate Fermi gas can be approximated
by the Sommerfeld expansion (e.g. see \cite{HuangStatMech}). Of particular
interest is the expression for entropy $S=\frac{\pi^{2}}{3}g(\epsilon_{F})k_{B}^{2}T,$
valid for $T\ll T_{F}$, where $g(\epsilon_{F})$ is the density of
states evaluated at the Fermi energy, $\epsilon_{F}$. For many applications
to Fermi gas experiments, the parameter of most interest is the degeneracy
parameter $t\equiv k_{B}T/\epsilon_{F},$ i.e. the ratio of the temperature
to the Fermi temperature, for which the Sommerfeld expression can
be written as
\begin{equation}
S=\frac{\pi^{2}}{3}\epsilon_{F}g(\epsilon_{F})k_{B}t.\label{eq:S}\end{equation}

Here we consider the change in degeneracy temperature
of a Fermi gas as it is slowly loaded from a harmonic trap into the
combined harmonic-lattice potential, as is done in experiments. To
characterize this temperature change we assume that the loading is
\emph{isentropic} so that the initial entropy in the harmonic potential
(with initial degeneracy temperature $t_{i}$) is the same as the
final entropy when the system is in the combined harmonic-lattice
potential (with final degeneracy temperature $t_{f}$). Within the
validity regime of the Sommerfeld relation (\ref{eq:S}), the ratio
of these temperatures is given by\begin{equation}
\frac{t_{f}}{t_{i}}=\frac{\kappa_{h}(N)}{\kappa_{c}(N)},\qquad(t_i,t_f \ll 1)\label{eq:degtratio}\end{equation}
obtained by assuming that $S$ remains constant, where we have introduced
the dimensionless extensive parameter $\kappa_{x}(N)\equiv\epsilon_{F,x}g_{x}(\epsilon_{F,x})\quad(x=h,c),$
with $g_{h}(\epsilon)$ and $\epsilon_{F,h}$ the density of states
and Fermi energy for a harmonic trap respectively. We have chosen
to express $\kappa$ as a function of $N$ rather than $\epsilon_{F},$
since the number of atoms remains constant during the loading procedure,
whereas the Fermi energy may change significantly. 

For the purely harmonic trap $g_{h}(\epsilon)=\epsilon^{2}/(2\hbar^{3}\bar{\omega}^{3})$
and $\epsilon_{F,h}=\hbar\bar{\omega}(6N)^{1/3}$ (e.g. see \cite{Butts1997a}),
which give\begin{equation}
\kappa_{h}(N)=3N.\label{eq:kappah}\end{equation}
 For $N<N_{c},$ the excited band states do not contribute to $\kappa_{c}$
and using results (\ref{eq:AnalyticInHomogDOS}) and (\ref{eq:EFc1})
we obtain \begin{equation}
\kappa_{c}(N)=\frac{3}{2}N,\qquad N<N_{c}.\label{eq:kappac}\end{equation}

Thus for $N<N_{c}$ the ratio of degeneracy temperatures will increase
by a factor of two. Because $N_{c}$ is a monotonically increasing
function of $V_{0}$, it might be expected that for sufficiently deep
final lattice depth we will always obtain this factor of 2 increase
in the degeneracy temperature. However, $N_{c}$ decreases with increasing
$\bar{\omega}$, and since $\bar{\omega}$ may change with lattice
depth (e.g.  if the lattice is produced by focused lasers) a large final
lattice depth  may in fact lead to $N>N_{c}$. We also note that
when $N<N_{c}$ the ratio of the degeneracy temperatures (\ref{eq:degtratio})
is independent of the harmonic trap frequency, even if this parameter
changes during the loading %
\footnote{However, the absolute temperature does depend on the trap frequency.%
}. Similar conclusions to those presented in this section are given in Ref. \cite{Kohl2006a}.

\begin{widetext}
\subsection{Numerical results for isentropic loading of the combined harmonic-lattice}

\begin{figure}[!t]
\includegraphics[width=6in]{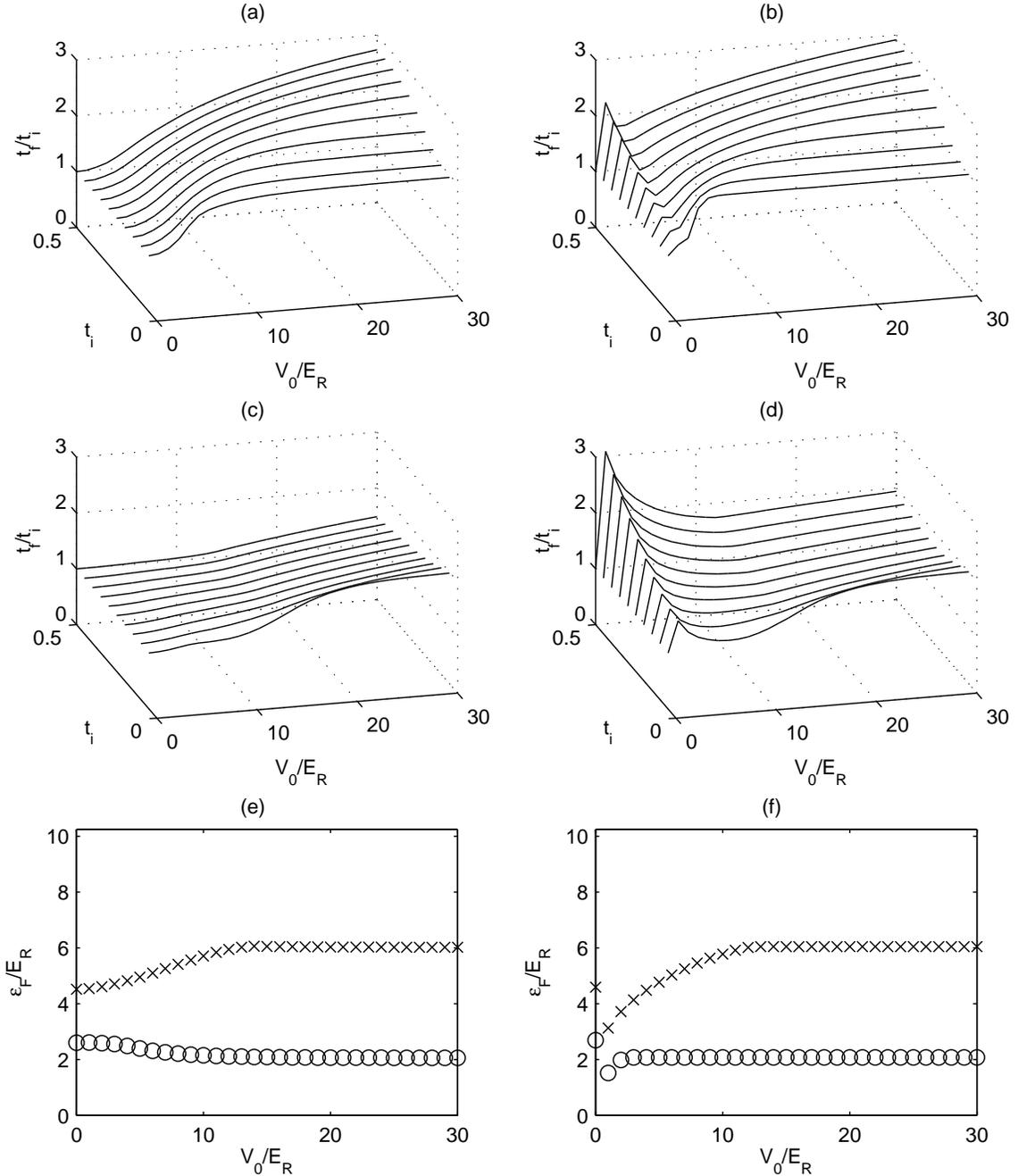}
\caption{\label{fig:mresults} Temperature of an ideal Fermi gas isentropically
loaded into a combined harmonic-lattice potential. The ratio of the
final degeneracy temperature (after loading) to the initial degeneracy
temperature in the harmonic trap are shown for various initial temperatures
and lattice depths for the case (a) $N=50\times10^{3}$ and (c) $N=250\times10^{3}$.
In (e) the Fermi energy is shown as a function of lattice depth for
$N=25\times10^{3}$ (circles), $N=250\times10^{3}$(crosses). (b), (d)
and (f) correspond to (a), (c) and (e) respectively, but are calculated using the
analytic density of states (\ref{eq:AnalyticInHomogDOS}). For all
results we have used an isotropic harmonic trap of frequency of $\bar{\omega}=0.04\omega_{R}.$ }
\end{figure} 
\end{widetext}

In this section we numerically examine the temperature of an ideal
Fermi gas loaded into a combined harmonic-lattice potential for a range of
lattice depths. Our main results are calculated using the energy spectrum
found by numerically diagonalizing (\ref{eq:Hsp}). To determine the
temperature under this type of loading we calculate the entropy of
a gas of $N$ fermions over a range of temperatures and lattice depths,
i.e. $S(T,V_{0},\vec{\omega},N)$, where $\vec{\omega}=\{\omega_{x},\omega_{y},\omega_{z}\}$
. We numercially invert this function to find temperature as a function
of the other quantities $T(S,V_{0},\vec{\omega},N)$, and by examining
the behaviour of $T$ for fixed $S$, we can predict the temperature
of the gas as a function of lattice depth.

Our procedure for determining entropy is as follows: The single particle
spectrum $\{\epsilon_{j}\}$ of the lattice is calculated for given
values of $\vec{\omega}$ and $V_{0}$. We then calculate the partition
function $\mathcal{Z}$\begin{equation}
\log\mathcal{Z}=\sum_{j}\log\left(1+e^{-\beta(\epsilon_{\mathbf{j}}-\mu)}\right),\label{eq:GrandPot}\end{equation}
 where $\mu$ is found by ensuring particle conservation. The entropy
of the system can then be expressed as\begin{equation}
S=k_{B}\left(\log\mathcal{Z}+\beta E-\mu\beta N\right),\label{eq:Entropy}\end{equation}
 where $\beta=1/k_{B}T$, and $E=-\partial\ln\mathcal{Z}/\partial\beta$
is the mean energy.

In Figs. \ref{fig:mresults}(a)-(f) we show the properties of an isentropically
loaded gas for various parameters. In Figs. \ref{fig:mresults}(a)
and (c) we show the ratio of the final to initial reduced temperatures
for $N=50\times10^{3}$ and $N=250\times10^{3}$ respectively. In
both cases, the reduced temperature is seen to increase as the lattice
depth increases. For $N=50\times10^{3}$ atoms the critical lattice
depth is $V_{c}\approx2.3E_{R}$. For the lowest initial reduced temperature
(i.e. the front-most curve in \ref{fig:mresults}(a)) we see that
the reduced temperature increases by a factor of two by the time that
$V_{0}$ increases beyond $V_{c}$ , in agreement with the predictions
of Eqs. (\ref{eq:degtratio})-(\ref{eq:kappac}) (also see Ref. \cite{Kohl2006a}).
Similarly, for the case of $N=250\times10^{3}$ atoms, $V_{c}\approx12.4E_{R}$
and for the lowest temperature result in Fig. \ref{fig:mresults}(c),
we see that the reduced temperature increases by a factor of two as
$V_{0}$ increases beyond this value of $V_{c}$. For higher initial
temperatures the Sommerfeld result does not hold. In Figs. \ref{fig:mresults}(a)
and (c) we see that the warmer systems (larger $t_{i}$ values) have
the contrasting behaviour of heating up more or less than the Sommerfeld
prediction respectively. In that regime the degeneracy temperature
is dominated by the change in the Fermi energy that occurs during
the lattice loading procedure, as shown in Fig. \ref{fig:mresults}(e).
That is, for the case in Fig. \ref{fig:mresults}(a) {[}\ref{fig:mresults}(c)]
the Fermi energy tends to  decrease [increase] with increasing lattice depth. 

In Fig. \ref{fig:mresults}(c) (and to a lesser extent in \ref{fig:mresults}(a))
we see that while excited bands are occupied (i.e. for $V<V_{c}$)
the degeneracy temperature increases by a factor of less than 2. This
suggests that having excited bands occupied might provide more favourable
conditions for investigating fermionic superfluidity in lattices.
Additionally, because the tunneling rate is larger for higher bands
it may be more easy to reversibly manipulate the lattice in this regime.

In Figs. \ref{fig:mresults}(b), (d), (f) we show the results equivalent
to those in Figs. \ref{fig:mresults}(a), (c), (e), but calculated
using the analytic density of states given in Eq. (\ref{eq:AnalyticInHomogDOS}).
Qualitatively the agreement between the results is good for $V_{0}\gtrsim4E_{R}$.
The main discrepancy is observed for small $V_{0}$ values where the
role of non-localized states and higher bands is important.

\section{Relation to experiments}

\begin{figure}[t]
\includegraphics[width=3.4in,keepaspectratio]{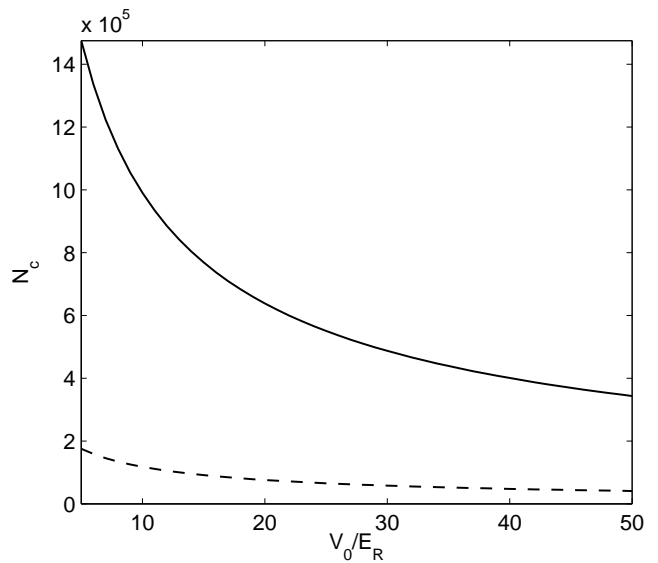}

\caption{\label{fig:NcExpt} $N_{c}$  for the case of $^{40}$K in a lattice
formed by focused lasers (solid) $\lambda=826$nm with $70\mu$m waist,
(dashed) $\lambda=1200$nm with $50\mu$m waist. }
\end{figure}

It is of interest to compare how important excited band effects might
be for current experiments. In Fig. \ref{fig:NcExpt} we show $N_{c}$
for parameters similar to those used in recent experiments by the
ETH group \cite{Kohl2005a,Stoferle2006a}. In those experiments up
to $10^{5}$ atoms were prepared in each spin state. Because the harmonic
confinement increases with lattice depth ($\bar\omega\sim\sqrt{V_{0}E_R}/\hbar$),
we find that the Fermi energy will eventually lie in the first excited  band, but only for very
large lattice depths ($V_{0}\gtrsim 280E_{R}$). We also consider a
longer wavelength lattice made from lasers with $\lambda=1200$nm
with slightly tighter focus (beam waist of $50\mu$m) for which the
recoil energy and the gap to higher bands is smaller. In such a configuration
(dashed curve) we see that higher bands would  become important for much
lower atom numbers.

\begin{figure}[t]
\includegraphics[width=3.4in,keepaspectratio]{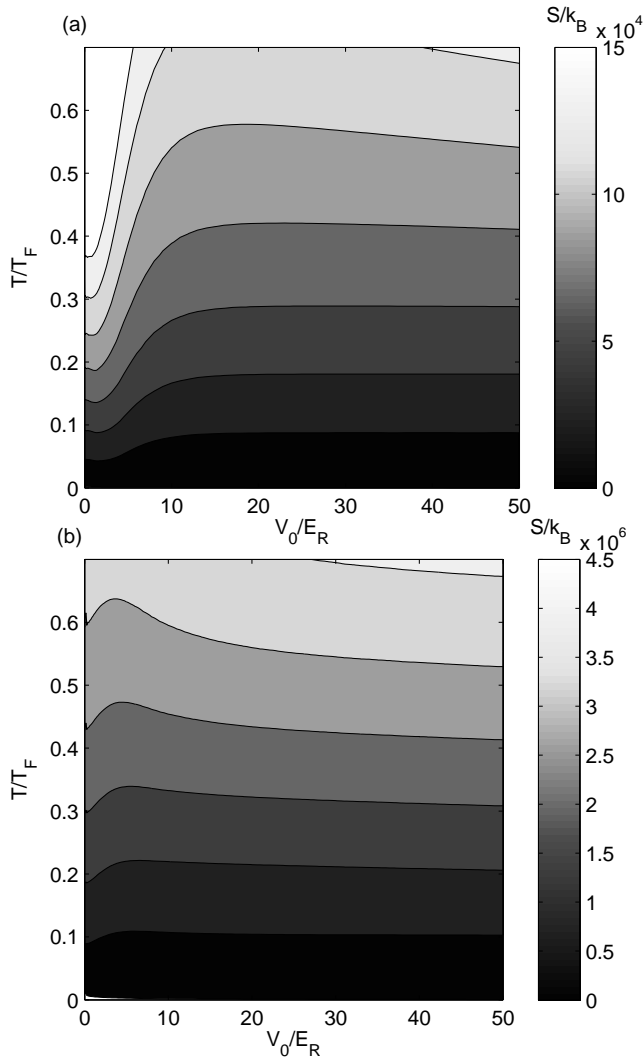}

\caption{\label{fig:Expt2} Degeneracy temperature for isentropic loading
of $^{40}$K into an optical lattice. (a) $N=50,000$ and (b) $N=750,000$.
Parameters correspond to solid line in \ref{fig:NcExpt}. Additional
harmonic confinerment of frequency $\omega=0.005\omega_{R}$ is superimposed
on the confinement due to the focused beam waist to make the spectrum
well-behaved as $V_{0}\to0$. }
\end{figure}

In Fig. \ref{fig:Expt2} we consider the effect of loading on the
degeneracy temperature for parameters relevant to the experiment in
Ref. \cite{Stoferle2006a}. The results in Fig. \ref{fig:Expt2}(a)
are for the same parameters considered by K\"ohl who made a calculation
in the tightbinding limit (see Fig. 1 of Ref. \cite{Kohl2006a}).
We broadly find agreement with those results, however make note of several differences.
First, at the lowest depths considered ($V_0\sim5E_R$)  K\"ohl   observed the reduced temperature  to initially decrease with increasing lattice depth. In contrast our results, which are valid for all lattice depths, do not exhibit this feature, and instead are seen to smoothly connect with  the $V_0=0E_R$ (purely harmonic case). We conclude that in this region the tightbinding approximation and the neglect of higher bands is not valid. 

Second, in  K\"ohl's results the reduced temperature is observed to saturate to a value of twice that of the initial harmonic trap  when the lattice is sufficiently deep (typically $V_0\gtrsim15E_R$). Many of our high temperature curves instead show a slight decrease in reduced temperature as the depth of the lattice increases. We have verified that this behaviour is due to higher band states, and that if we neglect them from our calculations our results saturate in agreement with those in \cite{Kohl2006a}. This indicates that even when the deep lattice behaviour is dominated by the ground band (i.e. we have $N<N_c$), if a  small number of atoms are able to thermally access excited band states they can have a significant effect on the temperature of the system during loading.

In Fig. \ref{fig:Expt2}(b) the same parameters are used, except the
number of atoms is increased to $750,000.$ According to Fig. \ref{fig:NcExpt}
for this number of atoms  $V_{c}\approx15E_{R}$.
In this case we see that only a small increase in the degeneracy temperature
occurs during loading and ultimately for sufficiently large final lattice depth (i.e. $V_0\gtrsim20E_R$) the reduced temperature is observed to be approximately the same as in the initial harmonic trap. We note that this result requires the occupation of higher
bands and cannot be analysed using a tightbinding approach.

The discussion in this section shows that while current experiments
are likely not strongly affected by higher bands, the parameter regime where
they become important is rather close. Experiments could enter this
regime by using larger wavelength lattices or tighter harmonic confinement
(e.g. more tightly focused lasers to produce the optical lattice).

\section{Conclusions}

We have discussed the nature of the spectrum of a Fermi gas in a combined
harmonic trap and optical lattice potential. Using this spectrum we
have derived an analytic density of states that is relatively accurate
for the lattice depths and harmonic confinements used in experiments.
We have characterized the validity criteria for this density of states
and have used it to characterise a Fermi gas in the combined potential.
As an application we have examined how adiabatic loading from a harmonic
trap into the combined harmonic-lattice potential affects the degeneracy
temperature of an ideal Fermi gas. Our results show that when excited band states are occupied
the system is less heated by the lattice loading, and may be less
sensitive to non-adiabatic effects, suggesting that this regime is
worthy of further investigation in experiments.

\section*{Acknowledgments}

PBB would like to acknowledge useful discussions with M. K\"ohl and
 support from the Marsden Fund of New Zealand. P.B. acknowledges a grant from the
{\it Lagrange Project}--CRT Foundation and is grateful
to the Jack Dodd Centre for the warm hospitality

\appendix

\section{Analytic approximation to band gap energy \label{sec:Egap}}
In this section we derive an analytic expression for the band gap in deep lattices. We go beyond the usual harmonic oscillator approximation and obtain results equivalent to those used in Ref. \cite{Spielman2006a}.

We will consider the standard harmonic approximation the optical
lattice potential, by making a Taylor expansion about the lattice
site minimum at $x=0$, i.e.
\begin{equation}
V_l=\frac{V_{0}}{2}[1-\cos(bx)]\approx\frac{V_{0}b^{2}}{4}x^{2}-\frac{V_{0}b^{4}}{48}x^{4},\label{eq:Vquartic}\end{equation}
where $b=2k$ is the reciprocal lattice vector.

Casting the harmonic term in the form of a harmonic oscillator
potential, \textbf{}$\frac{1}{2}m\omega_{\rm{Latt}}^{2}x^{2}$, yields the effective
harmonic oscillator frequency of $\omega_{\rm{Latt}}=\sqrt{V_{0}b^{2}/2m}$, and
the localized states in the optical lattice can be approximated as
harmonic oscillator states. This approximation neglects the influence
of tunneling between sites that gives a quasimomentum dependence,
and predicts that the band gap is equal to $\hbar\omega_{\rm{Latt}}$. For our
purposes it is desirable to go beyond this approximation and obtain
a more accurate analytic expression for the band gap energy. To do
this we use the harmonic oscillator states to treat the quartic term
in Eq. ( \ref{eq:Vquartic}) perturbatively. This is most easily done
using the normal ladder operators, so that $x^{4}=\left(\frac{\hbar}{2m\omega_{\rm{Latt}}}\right)^{2}(a^{\dagger}+a)^{4},$which
gives the 1st order shifts in the oscillator state energies as\begin{eqnarray}
\Delta E_{n}  & = & -\frac{E_{R}}{4}(2n^{2}+2n+1).\end{eqnarray} 

\begin{figure}[t]
\includegraphics[width=3.1in,keepaspectratio]{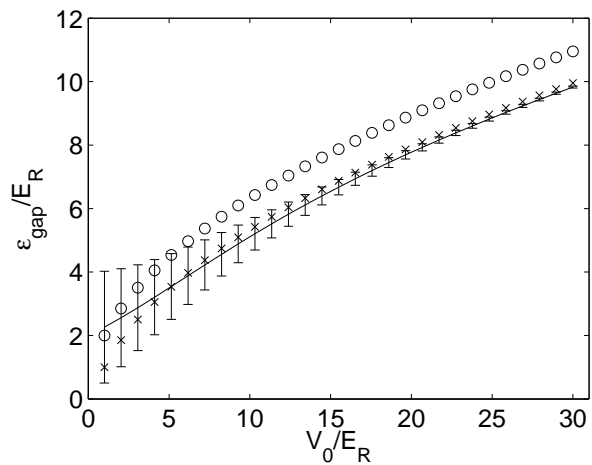}
\caption{\label{cap:Egap} Comparison between numerically determined energy
gap for a 1D sinusoidal lattice (solid), with the upper and lower
limits corresponding to the band gap at the center and the edge of
the Brillouin Zone. The harmonic approximation, $\hbar\omega_{{\rm Latt}}=2\sqrt{V_{0}E_{R}}$
(circles) and the estimate treating the quartic expansion perturbatively
, $\epsilon_{{\rm gap}}=2\sqrt{V_{0}E_{R}}-E_{R}$ (crosses).}
\end{figure}

Thus the approximate energies of the localized states are given by
\begin{equation}
E_{n}=\left[2\sqrt{\frac{V_{0}}{E_{R}}}(n+1/2)-\frac{1}{4}(2n^{2}+2n+1)\right]E_{R}.\end{equation}
Of most interest is the difference in energy between the $n=0$ and $n=1$  states, which provides an estimate of the energy gap between the ground and first excited bands, i.e. \begin{equation}
\epsilon_{\rm{gap}}= 2\sqrt{V_{0}E_R}-E_{R}.\end{equation}
We see that treating the quartic term leads to a 1 recoil suppression of the band gap compared to the harmonic oscillator frequency. In Fig. \ref{cap:Egap}
we compare this analytical expression with the band gap determined
numerically by evaluating the full band structure of the $\frac{V_{0}}{2}[1-\cos(bx)]$
potential. For shallow lattices the band gap is strongly dependent
on the value of quasimomenta considered and error bars indicate the
range of energy gaps from band centre to band edge. 

\bibliographystyle{apsrev}
\bibliography{Fermi_Bib}

\begin{thebibliography}{40}
\expandafter\ifx\csname natexlab\endcsname\relax\def\natexlab#1{#1}\fi
\expandafter\ifx\csname bibnamefont\endcsname\relax
  \def\bibnamefont#1{#1}\fi
\expandafter\ifx\csname bibfnamefont\endcsname\relax
  \def\bibfnamefont#1{#1}\fi
\expandafter\ifx\csname citenamefont\endcsname\relax
  \def\citenamefont#1{#1}\fi
\expandafter\ifx\csname url\endcsname\relax
  \def\url#1{\texttt{#1}}\fi
\expandafter\ifx\csname urlprefix\endcsname\relax\def\urlprefix{URL }\fi
\providecommand{\bibinfo}[2]{#2}
\providecommand{\eprint}[2][]{\url{#2}}

\bibitem[{\citenamefont{DeMarco and Jin}(1999)}]{DeMarco1999a}
\bibinfo{author}{\bibfnamefont{B.}~\bibnamefont{DeMarco}} \bibnamefont{and}
  \bibinfo{author}{\bibfnamefont{D.~S.} \bibnamefont{Jin}},
  \bibinfo{journal}{Science} \textbf{\bibinfo{volume}{285}},
  \bibinfo{pages}{1703} (\bibinfo{year}{1999}).

\bibitem[{\citenamefont{Schreck et~al.}(2001)\citenamefont{Schreck, Khaykovich,
  Corwin, Ferrari, Bourdel, Cubizolles, and Salomon}}]{Schreck2001a}
\bibinfo{author}{\bibfnamefont{F.}~\bibnamefont{Schreck}},
  \bibinfo{author}{\bibfnamefont{L.}~\bibnamefont{Khaykovich}},
  \bibinfo{author}{\bibfnamefont{K.~L.} \bibnamefont{Corwin}},
  \bibinfo{author}{\bibfnamefont{G.}~\bibnamefont{Ferrari}},
  \bibinfo{author}{\bibfnamefont{T.}~\bibnamefont{Bourdel}},
  \bibinfo{author}{\bibfnamefont{J.}~\bibnamefont{Cubizolles}},
  \bibnamefont{and} \bibinfo{author}{\bibfnamefont{C.}~\bibnamefont{Salomon}},
  \bibinfo{journal}{Phys. Rev. Lett.} \textbf{\bibinfo{volume}{87}},
  \bibinfo{pages}{080403} (\bibinfo{year}{2001}).

\bibitem[{\citenamefont{O'Hara et~al.}(2002)\citenamefont{O'Hara, Hemmer, Gehm,
  Granade, and Thomas}}]{O'Hara2002a}
\bibinfo{author}{\bibfnamefont{K.~M.} \bibnamefont{O'Hara}},
  \bibinfo{author}{\bibfnamefont{S.~L.} \bibnamefont{Hemmer}},
  \bibinfo{author}{\bibfnamefont{M.~E.} \bibnamefont{Gehm}},
  \bibinfo{author}{\bibfnamefont{S.~R.} \bibnamefont{Granade}},
  \bibnamefont{and} \bibinfo{author}{\bibfnamefont{J.~E.}
  \bibnamefont{Thomas}}, \bibinfo{journal}{Science}
  \textbf{\bibinfo{volume}{298}}, \bibinfo{pages}{2179} (\bibinfo{year}{2002}).

\bibitem[{\citenamefont{Modugno et~al.}(2002)\citenamefont{Modugno, Roati,
  Riboli, Ferlaino, Brecha, and Inguscio}}]{Modugno2002a}
\bibinfo{author}{\bibfnamefont{G.}~\bibnamefont{Modugno}},
  \bibinfo{author}{\bibfnamefont{G.}~\bibnamefont{Roati}},
  \bibinfo{author}{\bibfnamefont{F.}~\bibnamefont{Riboli}},
  \bibinfo{author}{\bibfnamefont{F.}~\bibnamefont{Ferlaino}},
  \bibinfo{author}{\bibfnamefont{R.~J.} \bibnamefont{Brecha}},
  \bibnamefont{and} \bibinfo{author}{\bibfnamefont{M.}~\bibnamefont{Inguscio}},
  \bibinfo{journal}{Science} \textbf{\bibinfo{volume}{297}},
  \bibinfo{pages}{2240} (\bibinfo{year}{2002}).

\bibitem[{\citenamefont{Gupta et~al.}(2003)\citenamefont{Gupta, Hadzibabic,
  Zwierlein, Stan, Dieckmann, Schunck, van Kempen, Verhaar, and
  Ketterle}}]{Gupta2003a}
\bibinfo{author}{\bibfnamefont{S.}~\bibnamefont{Gupta}},
  \bibinfo{author}{\bibfnamefont{Z.}~\bibnamefont{Hadzibabic}},
  \bibinfo{author}{\bibfnamefont{M.~W.} \bibnamefont{Zwierlein}},
  \bibinfo{author}{\bibfnamefont{C.~A.} \bibnamefont{Stan}},
  \bibinfo{author}{\bibfnamefont{K.}~\bibnamefont{Dieckmann}},
  \bibinfo{author}{\bibfnamefont{C.~H.} \bibnamefont{Schunck}},
  \bibinfo{author}{\bibfnamefont{E.~G.~M.} \bibnamefont{van Kempen}},
  \bibinfo{author}{\bibfnamefont{B.~J.} \bibnamefont{Verhaar}},
  \bibnamefont{and} \bibinfo{author}{\bibfnamefont{W.}~\bibnamefont{Ketterle}},
  \bibinfo{journal}{Science} \textbf{\bibinfo{volume}{300}},
  \bibinfo{pages}{1723} (\bibinfo{year}{2003}).

\bibitem[{\citenamefont{Regal et~al.}(2003)\citenamefont{Regal, Ticknor, Bohn,
  and Jin}}]{Regal2003a}
\bibinfo{author}{\bibfnamefont{C.~A.} \bibnamefont{Regal}},
  \bibinfo{author}{\bibfnamefont{C.}~\bibnamefont{Ticknor}},
  \bibinfo{author}{\bibfnamefont{J.~L.} \bibnamefont{Bohn}}, \bibnamefont{and}
  \bibinfo{author}{\bibfnamefont{D.~S.} \bibnamefont{Jin}},
  \bibinfo{journal}{Nature} \textbf{\bibinfo{volume}{424}}, \bibinfo{pages}{47}
  (\bibinfo{year}{2003}).

\bibitem[{\citenamefont{Cubizolles et~al.}(2003)\citenamefont{Cubizolles,
  Bourdel, Kokkelmans, Shlyapnikov, and Salomon}}]{Cubizolle2003a}
\bibinfo{author}{\bibfnamefont{J.}~\bibnamefont{Cubizolles}},
  \bibinfo{author}{\bibfnamefont{T.}~\bibnamefont{Bourdel}},
  \bibinfo{author}{\bibfnamefont{S.~J. J. M.~F.} \bibnamefont{Kokkelmans}},
  \bibinfo{author}{\bibfnamefont{G.}~\bibnamefont{Shlyapnikov}},
  \bibnamefont{and} \bibinfo{author}{\bibfnamefont{C.}~\bibnamefont{Salomon}},
  \bibinfo{journal}{Phys. Rev. Lett.} \textbf{\bibinfo{volume}{91}},
  \bibinfo{pages}{240401} (\bibinfo{year}{2003}).

\bibitem[{\citenamefont{Hofstetter et~al.}(2002)\citenamefont{Hofstetter,
  Cirac, Zoller, Demler, and Lukin}}]{Hofstetter2002}
\bibinfo{author}{\bibfnamefont{W.}~\bibnamefont{Hofstetter}},
  \bibinfo{author}{\bibfnamefont{J.~I.} \bibnamefont{Cirac}},
  \bibinfo{author}{\bibfnamefont{P.}~\bibnamefont{Zoller}},
  \bibinfo{author}{\bibfnamefont{E.}~\bibnamefont{Demler}}, \bibnamefont{and}
  \bibinfo{author}{\bibfnamefont{M.~D.} \bibnamefont{Lukin}},
  \bibinfo{journal}{Phys. Rev. Lett.} \textbf{\bibinfo{volume}{89}},
  \bibinfo{pages}{220407} (\bibinfo{year}{2002}).

\bibitem[{\citenamefont{Rabl et~al.}(2003)\citenamefont{Rabl, Daley, Fedichev,
  Cirac, and Zoller}}]{Rabl2003}
\bibinfo{author}{\bibfnamefont{P.}~\bibnamefont{Rabl}},
  \bibinfo{author}{\bibfnamefont{A.~J.} \bibnamefont{Daley}},
  \bibinfo{author}{\bibfnamefont{P.~O.} \bibnamefont{Fedichev}},
  \bibinfo{author}{\bibfnamefont{J.~I.} \bibnamefont{Cirac}}, \bibnamefont{and}
  \bibinfo{author}{\bibfnamefont{P.}~\bibnamefont{Zoller}},
  \bibinfo{journal}{Phys. Rev. Lett.} \textbf{\bibinfo{volume}{91}},
  \bibinfo{pages}{110403} (\bibinfo{year}{2003}).

\bibitem[{\citenamefont{Santos et~al.}(2004)\citenamefont{Santos, Baranov,
  Cirac, Everts, Fehrmann, and Lewenstein}}]{Santos2004}
\bibinfo{author}{\bibfnamefont{L.}~\bibnamefont{Santos}},
  \bibinfo{author}{\bibfnamefont{M.~A.} \bibnamefont{Baranov}},
  \bibinfo{author}{\bibfnamefont{J.~I.} \bibnamefont{Cirac}},
  \bibinfo{author}{\bibfnamefont{H.-U.} \bibnamefont{Everts}},
  \bibinfo{author}{\bibfnamefont{H.}~\bibnamefont{Fehrmann}}, \bibnamefont{and}
  \bibinfo{author}{\bibfnamefont{M.}~\bibnamefont{Lewenstein}},
  \bibinfo{journal}{Phys. Rev. Lett.} \textbf{\bibinfo{volume}{93}},
  \bibinfo{pages}{030601} (\bibinfo{year}{2004}).

\bibitem[{\citenamefont{Rigol and Muramatsu}(2004{\natexlab{a}})}]{Rigol2004a}
\bibinfo{author}{\bibfnamefont{M.}~\bibnamefont{Rigol}} \bibnamefont{and}
  \bibinfo{author}{\bibfnamefont{A.}~\bibnamefont{Muramatsu}},
  \bibinfo{journal}{Phys. Rev. A} \textbf{\bibinfo{volume}{69}},
  \bibinfo{pages}{053612} (\bibinfo{year}{2004}{\natexlab{a}}).

\bibitem[{\citenamefont{Viverit et~al.}(2004)\citenamefont{Viverit, Menotti,
  Calarco, and Smerzi}}]{viverit2004a}
\bibinfo{author}{\bibfnamefont{L.}~\bibnamefont{Viverit}},
  \bibinfo{author}{\bibfnamefont{C.}~\bibnamefont{Menotti}},
  \bibinfo{author}{\bibfnamefont{T.}~\bibnamefont{Calarco}}, \bibnamefont{and}
  \bibinfo{author}{\bibfnamefont{A.}~\bibnamefont{Smerzi}},
  \bibinfo{journal}{Phys. Rev. Lett.} \textbf{\bibinfo{volume}{93}},
  \bibinfo{pages}{110401} (\bibinfo{year}{2004}).

\bibitem[{\citenamefont{Modugno et~al.}(2003)\citenamefont{Modugno, Ferlaino,
  Heidemann, Roati, and Inguscio}}]{Modugno2003a}
\bibinfo{author}{\bibfnamefont{G.}~\bibnamefont{Modugno}},
  \bibinfo{author}{\bibfnamefont{F.}~\bibnamefont{Ferlaino}},
  \bibinfo{author}{\bibfnamefont{R.}~\bibnamefont{Heidemann}},
  \bibinfo{author}{\bibfnamefont{G.}~\bibnamefont{Roati}}, \bibnamefont{and}
  \bibinfo{author}{\bibfnamefont{M.}~\bibnamefont{Inguscio}},
  \bibinfo{journal}{Phys. Rev. A} \textbf{\bibinfo{volume}{68}},
  \bibinfo{pages}{011601(R)} (\bibinfo{year}{2003}).

\bibitem[{\citenamefont{Ott et~al.}(2004{\natexlab{a}})\citenamefont{Ott,
  de~Mirandes, Ferlaino, Roati, Modugno, and Inguscio}}]{Ott2004a}
\bibinfo{author}{\bibfnamefont{H.}~\bibnamefont{Ott}},
  \bibinfo{author}{\bibfnamefont{E.}~\bibnamefont{de~Mirandes}},
  \bibinfo{author}{\bibfnamefont{F.}~\bibnamefont{Ferlaino}},
  \bibinfo{author}{\bibfnamefont{G.}~\bibnamefont{Roati}},
  \bibinfo{author}{\bibfnamefont{G.}~\bibnamefont{Modugno}}, \bibnamefont{and}
  \bibinfo{author}{\bibfnamefont{M.}~\bibnamefont{Inguscio}},
  \bibinfo{journal}{Phys. Rev. Lett.} \textbf{\bibinfo{volume}{92}},
  \bibinfo{pages}{160601} (\bibinfo{year}{2004}{\natexlab{a}}).

\bibitem[{\citenamefont{St{\"o}ferle et~al.}(2006)\citenamefont{St{\"o}ferle,
  Moritz, G{\"u}nter, K{\"o}hl, and Esslinger}}]{Stoferle2006a}
\bibinfo{author}{\bibfnamefont{T.}~\bibnamefont{St{\"o}ferle}},
  \bibinfo{author}{\bibfnamefont{H.}~\bibnamefont{Moritz}},
  \bibinfo{author}{\bibfnamefont{K.}~\bibnamefont{G{\"u}nter}},
  \bibinfo{author}{\bibfnamefont{M.}~\bibnamefont{K{\"o}hl}}, \bibnamefont{and}
  \bibinfo{author}{\bibfnamefont{T.}~\bibnamefont{Esslinger}},
  \bibinfo{journal}{Phys. Rev. Lett.} \textbf{\bibinfo{volume}{96}},
  \bibinfo{pages}{030401} (\bibinfo{year}{2006}).

\bibitem[{\citenamefont{K{\"o}hl et~al.}(2005)\citenamefont{K{\"o}hl, Moritz,
  St{\"o}ferle, G{\"u}nter, and Esslinger}}]{Kohl2005a}
\bibinfo{author}{\bibfnamefont{M.}~\bibnamefont{K{\"o}hl}},
  \bibinfo{author}{\bibfnamefont{H.}~\bibnamefont{Moritz}},
  \bibinfo{author}{\bibfnamefont{T.}~\bibnamefont{St{\"o}ferle}},
  \bibinfo{author}{\bibfnamefont{K.}~\bibnamefont{G{\"u}nter}},
  \bibnamefont{and}
  \bibinfo{author}{\bibfnamefont{T.}~\bibnamefont{Esslinger}},
  \bibinfo{journal}{Phys. Rev. Lett.} \textbf{\bibinfo{volume}{94}},
  \bibinfo{pages}{080403} (\bibinfo{year}{2005}).

\bibitem[{\citenamefont{Ospelkaus et~al.}(2006)\citenamefont{Ospelkaus,
  Ospelkaus, Wille, Succo, Ernst, Sengstock, and Bongs}}]{Ospelkaus2006a}
\bibinfo{author}{\bibfnamefont{S.}~\bibnamefont{Ospelkaus}},
  \bibinfo{author}{\bibfnamefont{C.}~\bibnamefont{Ospelkaus}},
  \bibinfo{author}{\bibfnamefont{O.}~\bibnamefont{Wille}},
  \bibinfo{author}{\bibfnamefont{M.}~\bibnamefont{Succo}},
  \bibinfo{author}{\bibfnamefont{P.}~\bibnamefont{Ernst}},
  \bibinfo{author}{\bibfnamefont{K.}~\bibnamefont{Sengstock}},
  \bibnamefont{and} \bibinfo{author}{\bibfnamefont{K.}~\bibnamefont{Bongs}},
  \bibinfo{journal}{Phys. Rev. Lett.} \textbf{\bibinfo{volume}{96}},
  \bibinfo{pages}{180403} (\bibinfo{year}{2006}).

\bibitem[{\citenamefont{Hooley and Quintanila}(2004)}]{Hooley2004a}
\bibinfo{author}{\bibfnamefont{C.}~\bibnamefont{Hooley}} \bibnamefont{and}
  \bibinfo{author}{\bibfnamefont{J.}~\bibnamefont{Quintanila}},
  \bibinfo{journal}{Phys. Rev. Lett.} \textbf{\bibinfo{volume}{93}},
  \bibinfo{pages}{080404} (\bibinfo{year}{2004}).

\bibitem[{\citenamefont{Rigol and Muramatsu}(2004{\natexlab{b}})}]{Rigol2004b}
\bibinfo{author}{\bibfnamefont{M.}~\bibnamefont{Rigol}} \bibnamefont{and}
  \bibinfo{author}{\bibfnamefont{A.}~\bibnamefont{Muramatsu}},
  \bibinfo{journal}{Phys. Rev. A} \textbf{\bibinfo{volume}{70}},
  \bibinfo{pages}{043627} (\bibinfo{year}{2004}{\natexlab{b}}).

\bibitem[{\citenamefont{Rey et~al.}(2005{\natexlab{a}})\citenamefont{Rey,
  Pupillo, Clark, and Williams}}]{Rey2005a}
\bibinfo{author}{\bibfnamefont{A.~M.} \bibnamefont{Rey}},
  \bibinfo{author}{\bibfnamefont{G.}~\bibnamefont{Pupillo}},
  \bibinfo{author}{\bibfnamefont{C.~W.} \bibnamefont{Clark}}, \bibnamefont{and}
  \bibinfo{author}{\bibfnamefont{C.~J.} \bibnamefont{Williams}},
  \bibinfo{journal}{Phys. Rev. A} \textbf{\bibinfo{volume}{72}},
  \bibinfo{pages}{033616} (\bibinfo{year}{2005}{\natexlab{a}}).

\bibitem[{\citenamefont{Polkovnikov et~al.}(2002)\citenamefont{Polkovnikov,
  Sachdev, and Girvin}}]{Polkovnikov2002a}
\bibinfo{author}{\bibfnamefont{A.}~\bibnamefont{Polkovnikov}},
  \bibinfo{author}{\bibfnamefont{S.}~\bibnamefont{Sachdev}}, \bibnamefont{and}
  \bibinfo{author}{\bibfnamefont{S.~M.} \bibnamefont{Girvin}},
  \bibinfo{journal}{Phys. Rev. A} \textbf{\bibinfo{volume}{66}},
  \bibinfo{pages}{053607} (\bibinfo{year}{2002}).

\bibitem[{\citenamefont{Ruuska and T{\"o}rm{\"a}}(2004)}]{Ruuska2004a}
\bibinfo{author}{\bibfnamefont{V.}~\bibnamefont{Ruuska}} \bibnamefont{and}
  \bibinfo{author}{\bibfnamefont{P.}~\bibnamefont{T{\"o}rm{\"a}}},
  \bibinfo{journal}{New J. Phys.} \textbf{\bibinfo{volume}{6}},
  \bibinfo{pages}{59} (\bibinfo{year}{2004}).

\bibitem[{\citenamefont{Jaksch et~al.}(1998)\citenamefont{Jaksch, Bruder,
  Cirac, Gardiner, and Zoller}}]{Jaksch1998a}
\bibinfo{author}{\bibfnamefont{D.}~\bibnamefont{Jaksch}},
  \bibinfo{author}{\bibfnamefont{C.}~\bibnamefont{Bruder}},
  \bibinfo{author}{\bibfnamefont{J.~I.} \bibnamefont{Cirac}},
  \bibinfo{author}{\bibfnamefont{C.}~\bibnamefont{Gardiner}}, \bibnamefont{and}
  \bibinfo{author}{\bibfnamefont{P.}~\bibnamefont{Zoller}},
  \bibinfo{journal}{Phys. Rev. Lett.} \textbf{\bibinfo{volume}{81}},
  \bibinfo{pages}{3108} (\bibinfo{year}{1998}).

\bibitem[{\citenamefont{Buonsante et~al.}(2006)\citenamefont{Buonsante, Penna,
  Vezzani, and Blakie}}]{Buonsante2006a}
\bibinfo{author}{\bibfnamefont{P.}~\bibnamefont{Buonsante}},
  \bibinfo{author}{\bibfnamefont{V.}~\bibnamefont{Penna}},
  \bibinfo{author}{\bibfnamefont{A.}~\bibnamefont{Vezzani}}, \bibnamefont{and}
  \bibinfo{author}{\bibfnamefont{P.}~\bibnamefont{Blakie}}
  (\bibinfo{year}{2006}), \bibinfo{note}{arXiv:cond-mat/0610476}.

\bibitem[{\citenamefont{Ott et~al.}(2004{\natexlab{b}})\citenamefont{Ott,
  de~Mirandes, Ferlaino, Roati, T{\"u}rck, Modugno, and Inguscio}}]{Ott2004b}
\bibinfo{author}{\bibfnamefont{H.}~\bibnamefont{Ott}},
  \bibinfo{author}{\bibfnamefont{E.}~\bibnamefont{de~Mirandes}},
  \bibinfo{author}{\bibfnamefont{F.}~\bibnamefont{Ferlaino}},
  \bibinfo{author}{\bibfnamefont{G.}~\bibnamefont{Roati}},
  \bibinfo{author}{\bibfnamefont{V.}~\bibnamefont{T{\"u}rck}},
  \bibinfo{author}{\bibfnamefont{G.}~\bibnamefont{Modugno}}, \bibnamefont{and}
  \bibinfo{author}{\bibfnamefont{M.}~\bibnamefont{Inguscio}},
  \bibinfo{journal}{Phys. Rev. Lett.} \textbf{\bibinfo{volume}{92}},
  \bibinfo{pages}{120407} (\bibinfo{year}{2004}{\natexlab{b}}).

\bibitem[{\citenamefont{Kastberg et~al.}(1995)\citenamefont{Kastberg, Phillips,
  Rolston, Spreeuw, and Jessen}}]{Kastberg1995a}
\bibinfo{author}{\bibfnamefont{A.}~\bibnamefont{Kastberg}},
  \bibinfo{author}{\bibfnamefont{W.~D.} \bibnamefont{Phillips}},
  \bibinfo{author}{\bibfnamefont{S.~L.} \bibnamefont{Rolston}},
  \bibinfo{author}{\bibfnamefont{R.~J.~C.} \bibnamefont{Spreeuw}},
  \bibnamefont{and} \bibinfo{author}{\bibfnamefont{P.~S.}
  \bibnamefont{Jessen}}, \bibinfo{journal}{Phys. Rev. Lett.}
  \textbf{\bibinfo{volume}{74}}, \bibinfo{pages}{1542} (\bibinfo{year}{1995}).

\bibitem[{\citenamefont{Blakie and Porto}(2004)}]{Blakie2004b}
\bibinfo{author}{\bibfnamefont{P.~B.} \bibnamefont{Blakie}} \bibnamefont{and}
  \bibinfo{author}{\bibfnamefont{J.~V.} \bibnamefont{Porto}},
  \bibinfo{journal}{Phys. Rev. A} \textbf{\bibinfo{volume}{69}},
  \bibinfo{pages}{013603} (\bibinfo{year}{2004}).

\bibitem[{\citenamefont{Blakie and Bezett}(2005)}]{Blakie2005a}
\bibinfo{author}{\bibfnamefont{P.~B.} \bibnamefont{Blakie}} \bibnamefont{and}
  \bibinfo{author}{\bibfnamefont{A.}~\bibnamefont{Bezett}},
  \bibinfo{journal}{Phys. Rev. A} \textbf{\bibinfo{volume}{71}},
  \bibinfo{pages}{033616} (\bibinfo{year}{2005}).

\bibitem[{\citenamefont{Werner et~al.}(2005)\citenamefont{Werner, Parcollet,
  Georges, and Hassan}}]{Werner2005a}
\bibinfo{author}{\bibfnamefont{F.}~\bibnamefont{Werner}},
  \bibinfo{author}{\bibfnamefont{O.}~\bibnamefont{Parcollet}},
  \bibinfo{author}{\bibfnamefont{A.}~\bibnamefont{Georges}}, \bibnamefont{and}
  \bibinfo{author}{\bibfnamefont{S.~R.} \bibnamefont{Hassan}},
  \bibinfo{journal}{Phys. Rev. Lett.} \textbf{\bibinfo{volume}{95}},
  \bibinfo{pages}{056401} (\bibinfo{year}{2005}).

\bibitem[{\citenamefont{Rey et~al.}(2006)\citenamefont{Rey, Pupillo, and
  Porto}}]{Rey2006c}
\bibinfo{author}{\bibfnamefont{A.~M.} \bibnamefont{Rey}},
  \bibinfo{author}{\bibfnamefont{G.}~\bibnamefont{Pupillo}}, \bibnamefont{and}
  \bibinfo{author}{\bibfnamefont{J.~V.} \bibnamefont{Porto}},
  \bibinfo{journal}{Phys. Rev. A} \textbf{\bibinfo{volume}{73}},
  \bibinfo{pages}{023608} (\bibinfo{year}{2006}).

\bibitem[{\citenamefont{Ho and Zhou}(2007)}]{Ho2007a}
\bibinfo{author}{\bibfnamefont{T.-L.} \bibnamefont{Ho}} \bibnamefont{and}
  \bibinfo{author}{\bibfnamefont{Q.}~\bibnamefont{Zhou}}
  (\bibinfo{year}{2007}), \bibinfo{note}{arXiv:cond-mat/0703169}.

\bibitem[{\citenamefont{Blakie et~al.}(2007)\citenamefont{Blakie, Rey, and
  Bezett}}]{PBlakie2007a}
\bibinfo{author}{\bibfnamefont{P.~B.} \bibnamefont{Blakie}},
  \bibinfo{author}{\bibfnamefont{A.-M.} \bibnamefont{Rey}}, \bibnamefont{and}
  \bibinfo{author}{\bibfnamefont{A.}~\bibnamefont{Bezett}},
  \bibinfo{journal}{Laser Physics} \textbf{\bibinfo{volume}{17}},
  \bibinfo{pages}{198} (\bibinfo{year}{2007}).

\bibitem[{\citenamefont{K{\"o}hl}(2006)}]{Kohl2006a}
\bibinfo{author}{\bibfnamefont{M.}~\bibnamefont{K{\"o}hl}},
  \bibinfo{journal}{Phys. Rev. A} \textbf{\bibinfo{volume}{73}},
  \bibinfo{pages}{031601(R)} (\bibinfo{year}{2006}).

\bibitem[{\citenamefont{Rey et~al.}(2005{\natexlab{b}})\citenamefont{Rey,
  Blakie, Pupillo, Williams, and Clark}}]{Rey2005b}
\bibinfo{author}{\bibfnamefont{A.-M.} \bibnamefont{Rey}},
  \bibinfo{author}{\bibfnamefont{P.~B.} \bibnamefont{Blakie}},
  \bibinfo{author}{\bibfnamefont{G.}~\bibnamefont{Pupillo}},
  \bibinfo{author}{\bibfnamefont{C.~J.} \bibnamefont{Williams}},
  \bibnamefont{and} \bibinfo{author}{\bibfnamefont{C.~W.} \bibnamefont{Clark}},
  \bibinfo{journal}{Phys. Rev. A} \textbf{\bibinfo{volume}{72}},
  \bibinfo{pages}{023407} (\bibinfo{year}{2005}{\natexlab{b}}).

\bibitem[{\citenamefont{Blakie}(2006)}]{Blakie2006c}
\bibinfo{author}{\bibfnamefont{P.~B.} \bibnamefont{Blakie}},
  \bibinfo{journal}{New J. Phys.} \textbf{\bibinfo{volume}{8}},
  \bibinfo{pages}{157} (\bibinfo{year}{2006}).

\bibitem[{\citenamefont{Denschlag et~al.}(2002)\citenamefont{Denschlag,
  Simsarian, H\"{a}ffner, McKenzie, Browaeys, Cho, Helmerson, Rolston, and
  Phillips}}]{Denschlag2002a}
\bibinfo{author}{\bibfnamefont{J.~H.} \bibnamefont{Denschlag}},
  \bibinfo{author}{\bibfnamefont{J.~E.} \bibnamefont{Simsarian}},
  \bibinfo{author}{\bibfnamefont{H.}~\bibnamefont{H\"{a}ffner}},
  \bibinfo{author}{\bibfnamefont{C.}~\bibnamefont{McKenzie}},
  \bibinfo{author}{\bibfnamefont{A.}~\bibnamefont{Browaeys}},
  \bibinfo{author}{\bibfnamefont{D.}~\bibnamefont{Cho}},
  \bibinfo{author}{\bibfnamefont{K.}~\bibnamefont{Helmerson}},
  \bibinfo{author}{\bibfnamefont{S.~L.} \bibnamefont{Rolston}},
  \bibnamefont{and} \bibinfo{author}{\bibfnamefont{W.~D.}
  \bibnamefont{Phillips}}, \bibinfo{journal}{J. Phys. B}
  \textbf{\bibinfo{volume}{35}}, \bibinfo{pages}{3095} (\bibinfo{year}{2002}).

\bibitem[{\citenamefont{Huang}(1967)}]{HuangStatMech}
\bibinfo{author}{\bibfnamefont{K.}~\bibnamefont{Huang}},
  \emph{\bibinfo{title}{Statistical Mechanics}} (\bibinfo{publisher}{Wiley},
  \bibinfo{year}{1967}).

\bibitem[{\citenamefont{Butts and Rokhsar}(1997)}]{Butts1997a}
\bibinfo{author}{\bibfnamefont{D.}~\bibnamefont{Butts}} \bibnamefont{and}
  \bibinfo{author}{\bibfnamefont{D.}~\bibnamefont{Rokhsar}},
  \bibinfo{journal}{Phys. Rev. A} \textbf{\bibinfo{volume}{55}},
  \bibinfo{pages}{4346} (\bibinfo{year}{1997}).

\bibitem[{\citenamefont{Spielman et~al.}(2006)\citenamefont{Spielman, Johnson,
  Huckans, Fertig, Rolston, Phillips, and Porto}}]{Spielman2006a}
\bibinfo{author}{\bibfnamefont{I.}~\bibnamefont{Spielman}},
  \bibinfo{author}{\bibfnamefont{P.~R.} \bibnamefont{Johnson}},
  \bibinfo{author}{\bibfnamefont{J.}~\bibnamefont{Huckans}},
  \bibinfo{author}{\bibfnamefont{C.}~\bibnamefont{Fertig}},
  \bibinfo{author}{\bibfnamefont{S.}~\bibnamefont{Rolston}},
  \bibinfo{author}{\bibfnamefont{W.}~\bibnamefont{Phillips}}, \bibnamefont{and}
  \bibinfo{author}{\bibfnamefont{J.}~\bibnamefont{Porto}},
  \bibinfo{journal}{Phys. Rev. A} \textbf{\bibinfo{volume}{73}},
  \bibinfo{pages}{020702} (\bibinfo{year}{2006}).

\bibitem[{\citenamefont{Blakie and Clark}(2004)}]{Blakie2004a}
\bibinfo{author}{\bibfnamefont{P.~B.} \bibnamefont{Blakie}} \bibnamefont{and}
  \bibinfo{author}{\bibfnamefont{C.~W.} \bibnamefont{Clark}},
  \bibinfo{journal}{J. Phys. B} \textbf{\bibinfo{volume}{37}},
  \bibinfo{pages}{1391} (\bibinfo{year}{2004}).

\end{thebibliography}

\end{document}